\documentclass[apjl]{emulateapj}
\usepackage{natbib}
\usepackage{epsf}

\def\nG{\mbox{$\;\nu_{\rm GHz}$}}

\def\percc{\mbox{$\;{\rm cm}^{-3}$}}
\def\gmpercc{\mbox{$\;$gm cm$^{-3}$}}

\def\kms{\mbox{$\;$km s$^{-1}$}}     
\def\cms2{\mbox{$\;$cm s$^{-2}$}}    
\def\ergs{\mbox{$\;$ergs}}
\def\ergss{\mbox{$\;$ergs s$^{-1}$}}

\def\mug{\mbox{${\rm {\mu}G}$}}
\def\msun{\mbox{${{\rm M}_\odot}$}}

\def\arcsec{\rlap{$^{\prime\prime}$}\hbox to 2pt{}}
\def\arcmin{\rlap{$^{\prime}$}\hbox to 2pt{}}
\def\H0{\mbox{${\rm H_{0}}$}}
\def\mpc{\mbox{$\;{\rm Mpc}$}}
\def\myr{\mbox{$\;{\rm Myr}$}}

\def\mevcc{\mbox{$\;{\rm MeV\; {\rm cm}^{-3}}$}}

\def\kev{\mbox{$\;{\rm keV}$}}

\def\ev{\mbox{$\;{\rm eV}$}}

\def\yrs{\mbox{$\;{\rm yrs}$}}

\def\gtorder{\mathrel{\raise.3ex\hbox{$>$}\mkern-14mu
             \lower0.8ex\hbox{$\sim$}}}
\def\ltorder{\mathrel{\raise.3ex\hbox{$<$}\mkern-14mu
             \lower0.8ex\hbox{$\sim$}}}

\begin{document}
\bibliographystyle{plain}
\hyphenation{brems-strah-lung}

\title{Gamma Ray Emission from Merger Shocks in the Coma Cluster of Galaxies}
\author{Robert C. Berrington} 
\affil{University of Wyoming, P.\ O.\ Box 3905, Dept.\ of Physics \&
  Astronomy, Laramie, WY, 82071}
\email{rberring@uwyo.edu}
\author{Charles D.\ Dermer}
\affil{Naval Research Laboratory, Code 7653, 4555 Overlook SW, Washington, DC,
  20375-5352} 
\email{dermer@gamma.nrl.navy.mil}

\begin{abstract}
A numerical simulation model of the injection and cooling of nonthermal
particles energized by shocks formed in merging clusters of galaxies is used
to fit radio and X-ray data observed from the Coma cluster of galaxies. The
results are consistent with a primary merger-shock origin for both the diffuse
radio halo emission and the hard X-ray excess measured with {\it Beppo-SAX}
and {\it RXTE}. For equal (1\%) efficiency of power injected in nonthermal
protons and electrons, we predict that the Coma cluster of galaxies will be
significantly detected with the space-based observatory {\it GLAST}, and
marginally detectable with the ground-based $\gamma$-ray observatories {\it
VERITAS} and {\it HESS}.  Significant TeV detections are possible if the
nonthermal proton intensity is greater due to a larger efficiency of
nonthermal hadron acceleration, or to past merger events. The nonthermal
hadronic content in Coma is also revealed by a weak, hard secondary emission
component at $\sim 10$ -- 100 GHz. The spectral softening of the radio
emission at large radii from the Coma cluster core derives in this scenario
from the decreasing shock speed away from cluster center for an on-axis merger
event. We discuss differences between merger and delayed turbulence models.
\end{abstract}

\section{Introduction}
\label{sec:introduction}

The discovery of the first radio halo, Coma C, demonstrated the existence of a
population of nonthermal electrons within the intracluster medium (ICM) of the
Coma cluster of galaxies (see \citet{sarazin:88} and \cite{biviano:98} for
review; radio intensity maps and spectra of Coma are reported by
\citet{thierbach:03}, \citet{schlickeiser:87}, and \cite{giovannini:93}).  The
origin of these energetic electrons could be associated with a merger event
\citep{tribble:93}, for example, between the central NGC 4889 and NGC 4874
galaxy groups.  In addition, the radio relic 1253+275 is found on the Coma
cluster (A1656) periphery, possibly indicating the presence of a shock
resulting from a recent or ongoing merger event.  Several tens of galaxy
clusters, representing $\sim 10$\% of the cluster population, are reported to
have radio halos, and these clusters preferentially show recent merger
activity \citep{buote:01}. Because of their relatively short radiation
lifetimes, GHz-emitting nonthermal electrons have to be accelerated on
timescales shorter \citep{sarazin:99,petrosian:01} than merger timescales.

Optical measurements of radial velocities of galaxies reveal the cluster
dynamics dominated by the gravitational potential well of the dark-matter
halo.  Although Coma is considered to be the nearest rich relaxed cluster,
with a FWHM diffuse X-ray angular extent $\sim 30^\prime$, optical surveys of
the radial velocities of 1174 galaxies in the Coma cluster
\citep{colless:96,biviano:96,edwards:02} show that it is in a highly dynamic
state as a result of a recent merger event.

Observations with {\it ROSAT} \citep{white:93} and {\it ASCA}
(\citet{honda:96}; see references in \citep{briel:01}) show multiple X-ray
emission peaks from the core of the Coma cluster.  The diffuse cluster
emission peak associated with the cluster center is coincident with the galaxy
NGC 4874.  A second diffuse emission peak was found approximately halfway
between NGC 4874 and NGC 4889. {\it XMM-Newton} observations
\citep{arnaud:01,briel:01} did not detect this second peak but found X-ray
emission peaks coincident with both NGC 4874 and NGC 4889, a relaxed thermal
structure within 10$^\prime$ of the cluster center, multiple point sources,
and stripped gas on the cluster periphery. {\it Chandra} observations of
galaxy clusters with radio halos reveal complex temperature maps indicating
recent or ongoing merger events \citep{govoni:04}. In the case of Coma, {\it
Chandra} results indicate that the X-ray emission from the coronae surrounding
the dominant galaxies NGC 4874 and NGC 4889 and their galaxy groups arises
from denser, X-ray gas clouds with temperatures of a few keV that are in
equilibrium with the intracluster gas \citep{vikhlinin:01}.

{\it BeppoSAX} \citep{fusco-femiano:04} and Rossi X-Ray Timing Explorer ({\it
RXTE}) \citep{rephaeli:02} observations of the Coma Cluster show hard X-ray
(HXR) emission in excess of single-temperature bremsstrahlung, though its
detection with {\it BeppoSAX} is disputed \citep{rossetti:04}, and the hard
tail remains unconfirmed with {\it INTEGRAL}.  {\it BeppoSAX} results
\citep{nevalainen:04} also indicate that clusters undergoing recent or ongoing
merger events have HXR excesses with $20$ -- $80\kev$ luminosities $\approx
10^{43}$ -- $10^{44}\ergs$, whereas fully relaxed clusters are consistent with
no HXR excess.

At $\gamma$-ray energies, the Coma cluster was not detected \citep{reimer:03}
with the EGRET telescope on the {\it Compton Observatory} at $\gtrsim 100$
MeV.  Statistical analysis of the positional association of unresolved
gamma-ray emission from rich clusters of galaxies in the local universe shows
a weak ($\sim 3\sigma$) positive correlation \citep{scharf:02}. In addition,
there are claims of an association between unidentified, high galactic
latitude ($|b| > 20^{\circ}$) EGRET sources and galaxy clusters
(\citet{kawasaki:02,colafrancesco:02}, though contested by
\citet{reimer:03}). The results of searches for TeV emission with the
ground-based imaging air Cherenkov telescopes (IACTs) {\it HESS} (High Energy
Stereoscopic System) and {\it VERITAS} (Very Energetic Radiation Imaging
Telescope Array System) as they reach their design capabilities are anxiously
awaited,\footnote{The latitudes of Whipple/VERITAS and HESS are $+31.4^\circ$
and $-23.3^\circ$, respectively, so Coma, at declination $+38.0^\circ$, is a
better target for {\it VERITAS} than {\it HESS}.}  but the sensitivities of
IACTs and coded mask telescopes such as {\it INTEGRAL} are compromised by the
extent of Coma's diffuse emission \citep{gabici:04} which, as in the case of
the thermal X-rays, could subtend a large fraction of a square degree.  The
diffuse extent of the nonthermal emission in clusters of galaxies, as
indicated by radio halos and relics, also makes it more difficult to map
emission from cluster and accretion shocks with narrow field-of-view pointing
telescopes.  The $\gtrsim$ TeV $\gamma$-rays from Coma will also be attenuated
by the diffuse extragalactic infrared radiation \citep{gabici:04}.

There have been numerous attempts to model the diffuse radio emission
associated with cluster radio halos in the context of a cluster merger
model. \citet{miniati:01} treat acceleration of cosmic-ray protons in the
context of a cosmological structure-formation simulation.  They find that
strong shocks form from cool, low density ICM gas accreting onto the cluster
periphery.  Such shocks are not expected to penetrate into the central region
of the cluster environment.  Numerical simulations
\citep{roettiger:99,ricker:98,ricker:01} show that shocks forming in merger
events between two virialized clusters of galaxies can, however, penetrate
into the cluster core.

\citet{gabici:03} found that shocks resulting from a major merger event (as
defined in Section 2) between two virialized clusters of similar mass produce
nonthermal particle spectra that are too soft to account for the observed
emission from most radio halos. Minor mergers are more effective for injecting
hard electron spectra.  Berrington \& Dermer (2003) (hereafter
\citet{berrington:03}) show that the typical peak Mach speeds ${\cal M}$ of
cluster merger shocks in minor mergers are ${\cal M} \approx 3$ -- 5. Primary
electron synchrotron radiation injected by merger shocks can account for the
observed radio emission from Coma C, as we show here.

Besides a directly accelerated electron population from a merger shock,
nonthermal radio halos detected from Coma and more than $\sim 20$ clusters of
galaxies could also originate from the decay products of proton interactions
\citep{dennison:80, ves87}. \citet{blasi:99} performed a detailed study of
this model for the Coma cluster and find that if the radio emission is due to
secondary electrons, then EGRET should have detected Coma.  Moreover, this
model is difficult to reconcile with the correlation of cluster radio halos
with merger activity \citep{buote:01} or the softening of Coma's radio
spectrum with radius, because clusters of galaxies will confine protons for
the age of the universe \citep{volk:96,berezinsky:97}. The cluster radio halos
could also be a consequence of delayed acceleration by second-order Fermi
processes associated with magnetic turbulence in the ICM
\citep{schlickeiser:87}. The most detailed study to date \citep{brunetti:04}
of the plasma physical processes results in a model for the Coma radio
spectrum that could fit the data.  Some concerns with the merger model, and
fits to Coma data in a simplified merger model is given by \citet{blasi:01}
(with references to earlier work).  We consider these points in our Discussion
section.

In Section 2, we adapt our merger-shock model \citep{berrington:03} to the
Coma cluster environment. Our approach directly ties the substructure in Coma
to the radio and HXR excess.  Assuming a merger-shock origin for the
nonthermal radiation from the Coma cluster, we show in Section 3 that if
protons are accelerated with the same efficiency as electrons, Coma is
predicted to be detected at GeV energies with the Gamma Ray Large Area Space
Telescope ({\it GLAST}), and at TeV energies with the IACTs {\it VERITAS} and
{\it HESS}. Hadrons remaining from previous merger shocks \citep{gabici:03}
could increase the $\gamma$-ray flux, but loss of sensitivity to extended
sources could decrease detectability.  An estimate of GLAST sensitivity for
the Coma cluster in the Appendix shows that even for the most optimistic
fluxes of $\gamma$-rays allowed by radio observations, Coma will be detected
with only $\sim 5\sigma$ significance.

The results are summarized in Section 4, where we discuss and compare
predictions of merger and reacceleration models. We argue that gamma-ray
detection of Coma will support a merger-shock origin of the nonthermal
radiation from Coma, though second-order turbulent re-acceleration effects
could play an associated role in nonthermal emissions from clusters of
galaxies.  The Appendix gives some analytic results for cluster merger physics
that provide a check on the numerical results.

\section{The Coma Cluster Environment}

Because nonthermal spectral power depends sensitively on the parameters of the
cluster environment, we modified the code developed by \citet{berrington:03}
specifically to the properties of Coma (see Table 1).  The Coma cluster of
galaxies is well-described by an event where a dominant cluster of total mass
$M_{1} \cong 0.8 \times 10^{15} \msun$ \citep{colless:96} merges with a
smaller cluster of total mass $M_{2} \approx 0.1 \times 10^{15}\msun$ cluster
\citep{vikhlimin:97,donnelly:99}. We assume a gas mass fraction of $\approx
5\%$.

\begin{deluxetable}{ll}
\tablecolumns{2}
\tablewidth{0pt}
\tablecaption{Properties of Merging Clusters in Coma}
\tablehead{}
\startdata
Redshift $z$ & $0.023$\\
$d_L(\rm cm)$ & $3.0\times 10^{26}~{\rm cm}$\\
$M_1$ & $0.8\times 10^{15} M_\odot$ \\
$M_2$ & $0.1\times 10^{15} M_\odot$ \\
$\langle T_X \rangle_1 $ & 8.21 keV \\
$\langle T_X \rangle_2 $ & 2.0 keV \\
$r_{c,1}$ & 0.257 Mpc \\
$r_{c,2}$ & 0.15 Mpc \\
$r_{{\rm max},1}$ & 1.3 Mpc \\
$r_{{\rm max},2}$ & 0.75 Mpc \\
$\beta$ & 0.705 \\
$\rho_{0,1}$ & $6.39\times 10^{-27}$ gm cm$^{-3}$ \\
$\rho_{0,2}$ & $1.67\times 10^{-27}$ gm cm$^{-3}$ \\
\enddata
\end{deluxetable}

Let $\Delta_{m}$ represent the mass ratio between the merging and dominant
clusters. Merger events with mass ratio $\Delta_{m} \leq 0.1$ are designated
accretion events \citep{salvador-sole:98}.  The core of the dominant cluster
is found to remain intact unless $\Delta_{m} \geq 0.6$ \citep{fujita:01}.  In
our proposed model, the mass ratio of the two clusters is $\Delta_{m} \cong
0.13$.  This places our model well below the threshold when the internal
structure of the dominant core is destroyed, and classifies our model as a
cluster merger event.  In addition, the mass ratio is great enough that the
shock resulting from the merger event will penetrate into the core of the
cluster \citep{roettiger:99}.  Our semi-analytical dynamics approach compares
favorably with N-body simulation results until the centers-of-mass of the
clusters pass through each other \citep{berrington:03}. At later times, the
dynamics of the merging cold dark matter halos and the hydrodynamics of the
gas are not accurately treated, and during this phase we expect rapid
coalescence of the cold dark matter halo and shock quenching as the gas
dynamics enters a Sedov phase.

The calculation of nonthermal electron injection, electron energy losses and
the production of the Compton-scattered emission for the various radiation
fields, including the cosmic microwave background radiation (CMBR), the
stellar radiation, and the X-ray radiation fields, follows the treatment of
\citet{berrington:03} (see also \citet{sarazin:99,petrosian:01}).  For a
galaxy luminosity function described by the Schechter luminosity function with
parameters $M^{*} = -21.26$, $\xi_{*} = 107$, and $\alpha = 1.25$
\citep{schechter:76}, the energy density of the stellar radiation field is
found to be
\begin{equation} 
\rho_{\star} = 4.87 \times 10^{-9} \left( \frac{R}{1\mpc} \right)^{-2} \mevcc
\label{eqn:stellar_radiation_field}
\end{equation}
where R is the radius of the Coma cluster.  

The Coma cluster has a tenuous ICM that emits thermal X-ray bremsstrahlung.
To model the emission from the ICM, we assume that the ICM is well
approximated by a $\beta$ model\footnote{As shown in Appendix A.3, the use of
an NFW profile gives relative merger speeds that are similar to that obtained
with a $\beta$-model.} with core radius $r_{c} = 0.257 \mpc$, central electron
density $\rho_{e0} = 3.82 \times 10^{-3}\percc$, central proton density
$\rho_{0,1} = 6.39 \times 10^{-27} \gmpercc$, power-law slope $\beta = 0.705$,
and a mean gas temperature $\langle T_{\rm X} \rangle = 8.21\kev$
\citep{mohr:99} (for the total mass modeling, we assume for simplicity that
the gas is entirely composed of H).  The energy density of the X-ray radiation
field is then given by
\begin{equation} 
\rho_{X} = 1.30 \times 10^{-10} \left( \frac{R}{1\mpc} \right)^{-2}\mevcc.
\label{eqn:x-ray_radiation_field}
\end{equation}
This energy density corresponds to a total free-free luminosity of $\sim 1
\times 10^{45}\ergss$ emitted within a cluster radius of $1\mpc$.  The energy
density of the CMBR is
\begin{equation} 
\rho_{\rm CMB} = 2.7 \times 10^{-7}\mevcc.
\label{eqn:CMBR}
\end{equation}

Given the evidence for a recent merger event in Coma, as described in Section
\ref{sec:introduction}, we argue that merger shocks accelerate nonthermal
particles to produce nonthermal radio and X-ray emission.  The evolution and
properties of the forward and reverse cluster merger shocks are determined by
the relative speeds and densities of the merging clusters.  We have calculated
the relative cluster speeds and shock properties following the method
described in \citet{berrington:03}.  The particle number density of the ICM is
assumed to be well approximated by a spherically-symmetric $\beta$ model whose
parameters are given above.  The dark matter halo, whose total mass is
normalized to $M_{1}$ out to a maximum radius $r_{\rm max}$ as given in Table
1, is assumed to follow a profile similar to the cluster ICM. {\it Beppo-SAX}
observations \citep{ettori:02} support the claim that the dark matter halo of
Coma is better fit by the \citet{king:62} approximation to the isothermal
model, given by
\begin{equation}
\rho(r) = \rho_0\left[1 + \left(\frac{r}{r_{c}}\right)^{2}\right]^{-3/2}\;,
\label{eqn:king_approximation}
\end{equation}
where $\rho_0 = \zeta \rho_{0,1}$ is the central density, $\zeta$ is the ratio
of the total matter density to the normal matter density $\rho_{0,1}$ and
$r_{c}$ is the core radius, than the \citet{navarro:97} (NFW) density profile.

The efficiency parameters $\eta_{e}$ and $\eta_{p}$ giving the fraction of
power dissipated by the shock in the form of nonthermal electrons and protons,
respectively, are set equal to 1\%. The injected particle spectrum is
described by a power law in momentum space truncated by an exponential
cutoff. The cutoff energy is determined by time available for acceleration
since the onset of the shock, comparison of the size scale of the system with
the particle Larmor radius, and competition of the acceleration time scale
with radiative-loss time scales (\S2.2 in \citet{berrington:03}). The total
energy liberated from the cluster merger event and deposited into the
nonthermal particles is $\sim 1.5 \times 10^{61}\ergs$.

\section{Model Results}

The model that best represents the multiwavelength data of the Coma Cluster is
found to be described by the merger of two clusters nearing collision time,
$t_{\rm coll}$, when the centers of mass of the two clusters are almost
coincident, as supported by observations
\citep{colless:96,biviano:96,edwards:02}. With the masses of the merging
clusters given above, and an initial separation of 2.25 Mpc, the observations
take place at $t_{\rm coll} \approx 1.0 \times 10^{9}\yrs$ after the onset of
the merger.  The observed redshift of the Coma cluster is $z_{o} = 0.023$.
For this value of $t_{\rm coll}$, the redshift of the cluster at the start of
the cluster merger event (when the outer radii of the two $\beta$-model
profiles first intersect) for a flat $\Lambda$CDM cosmology with $\Omega_{0} =
0.3$ and $\Omega_{\Lambda} = 0.7$ is $z_{i} = 0.10$.  Here we use a Hubble
constant $H_{0} = 75$ km s$^{-1}$ Mpc$^{-1}$, giving a luminosity distance
$d_L = 3.0\times 10^{26}$ cm for Coma.

\begin{figure}[t]
\vskip-1.0in
\begin{center}
\leavevmode
\hbox{%
\epsfxsize=3.0in
\epsffile{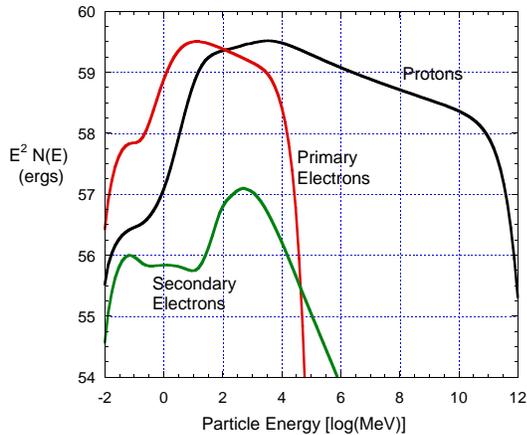}}
\vskip-0.5in
\caption{Total energy spectra of primary protons and primary and secondary
electrons, plotted in the form of $E^2 N(E)$, used to fit the measured
spectral energy distribution of the Coma cluster.  }
\label{fig1}
\end{center}
\end{figure}

The total energy spectra of primary protons and primary and secondary
electrons (including positrons) that fit measurements of the Coma spectrum are
shown in Fig.\ \ref{fig1}. As can be seen, the proton distribution is well
described by a power-law spectrum $N(E)\propto E^{-2.2}$ between $\approx 1$
GeV and $\approx 10^{17}$ eV. At the highest energies, the spectrum is cut off
due to the available time required to accelerate protons to the highest
energies. The strong radiative losses, principally due to Compton scattering
with the CMBR, causes the abrupt cutoff of the primary electron
spectrum. Because of the production kinematics and pile-up due to radiative
losses, the secondary electron spectrum displays a bump at a few GeV.

\begin{figure}[t]
\vskip-1.0in
\begin{center}
\leavevmode
\hbox{%
\epsfxsize=3.0in
\epsffile{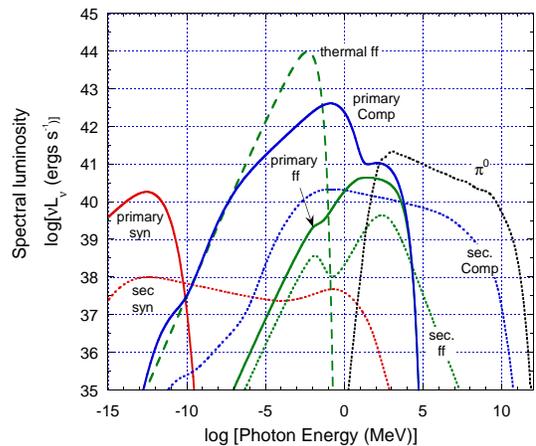}}
\caption{Spectral emission components for the fit to the volume-integrated
multiwavelength spectral energy distribution observed from the Coma cluster of
galaxies. Synchrotron, Compton, and bremsstrahlung emission components are
shown for the directly-accelerated primary nonthermal electrons (solid curves)
and for nonthermal leptons and $\pi^0$-decay $\gamma$ rays formed as
secondaries in inelastic nuclear pion-producing processes (dotted
curves). Also shown is the diffuse thermal bremsstrahlung emission (dashed
curve).  }
\label{fig2}
\end{center}
\end{figure}

Fig.\ \ref{fig2} shows the different spectral components used to fit the
multiwavelength Coma spectrum.  For the single merger event calculated here,
the primary electron radiation generally dominates secondary emissions at
radio and X-ray wavelengths, but secondary emissions make a contribution at mm
radio wavelengths and at $\gtrsim 100$ MeV energies for the assumed 1\%
efficiency fractions $\eta_e$ and $\eta_p$ of energy injected in the form of
nonthermal particles. The value of $\eta_e$ is determined by normalizing the
primary electron Compton emissions to the $\gtrsim 40$ keV nonthermal X-ray
flux, insofar as only a small fraction of the X-ray flux is produced by
bremsstrahlung and Compton emissions from secondary $\pi^\pm \rightarrow
e^\pm$ processes.  The hardening at $\approx 1$ GeV in the primary Compton
component is stellar radiation photons scattered by the primary electrons. The
ratio of the energy densities, and therefore peak fluxes, is $\approx 50$,
from eqs.\ (\ref{eqn:stellar_radiation_field}) and
(\ref{eqn:CMBR}). Scattering of the thermal diffuse X-rays to $>$ GeV energies
is suppressed by the absence of primary electrons at these energies (Fig.\
\ref{fig1}).

\begin{figure}[t]
\vskip-1.0in
\begin{center}
\leavevmode
\hbox{%
\epsfxsize=3.0in
\epsffile{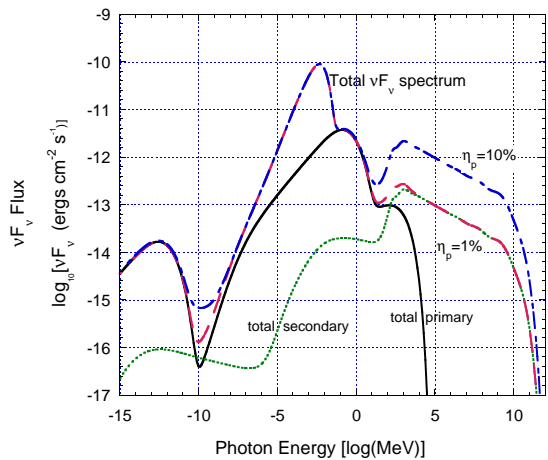}}
\caption{Total spectral energy distribution used to fit observations of the
Coma cluster with 1\% of the power swept up by the forward and reverse shocks
dissipated in nonthermal electrons. The two curves with $\eta_p = 1$\% and
$\eta_p = 10$\% correspond to the fraction of shock power deposited in
nonthermal protons. Separate components for the total primary (solid curve)
and total secondary (dotted curve) emissions for the case $\eta_p = 1$\% are
also shown.}
\label{fig3}
\end{center}
\end{figure}

At $\gtrsim 100$ MeV $\gamma$-ray energies, most of the emission is from
secondary $\pi^0$-decay $\gamma$ rays formed by nuclear processes in
collisions of nonthermal protons with the diffuse gas. The flux of the
secondary emissions is in proportion to the assumed fraction, 1\%, of the
power swept-up by the forward and reverse shocks and injected in the form of a
nonthermal power-law momentum distributions of hadrons.  The total spectral
energy distribution used to fit the Coma spectrum is shown in Fig.\
\ref{fig3}, where we vary the fraction $\eta_p$ of energy deposited in the
form of nonthermal hadrons from $ 1$\% to $\eta_p = 10$\%.

\begin{figure}[t]
\begin{center}
\leavevmode
\hbox{%
\epsfxsize=3.0in
\epsffile{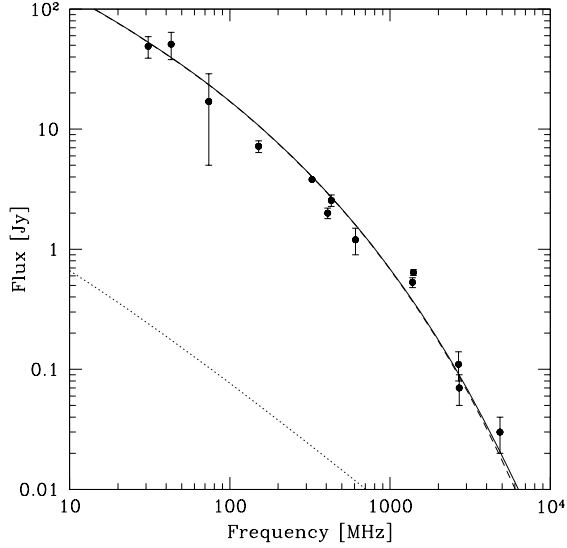}}
\caption{Comparison of radio observations of the Coma cluster with results of
the cluster merger-shock model.  The solid curve is the total radio emission
from the model, the dashed curve is the contribution to the total emission
from primary electrons, and the dotted curve from secondary electrons and
positrons.  The solid circles are observational data points
\citep{thierbach:03}. }
\label{fig:coma_radio}
\end{center}
\end{figure}

A comparison of the merger model with the measured \citep{thierbach:03} radio
emission from the radio halo of Coma is shown in Figure \ref{fig:coma_radio}.
The mean magnetic field, obtained by normalizing to the radio emission, is
$0.22\;\mug$, and is in agreement with analytic estimates (see Appendix).  Our
model is in accord with a primary electron source for the radio emission---a
comparable contribution to the radio flux from secondary nuclear production
would require that $\eta_p$ approach 100\% to produce the measured radio
emission, and would furthermore require a very different nonthermal proton
spectrum. From spectral considerations, observations of Coma imply that
$\eta_p \lesssim 20$\%; otherwise a hard hadronic component would be seen in
the multi-GHz radio emission. We note that the fluxes from secondary electrons
and positrons calculated here may overestimate the actual fluxes from
secondary leptons by as much as a factor of $\approx 2$ because a uniform
density profile, appropriate to the central region of the cluster, is used in
the calculation of secondary production. This is nevertheless a good
approximation because the density in the central region is roughly constant
and most of the nonthermal radiation is formed by interactions in this region.

\begin{figure}[t]
\begin{center}
\leavevmode
\hbox{%
\epsfxsize=3.0in
\epsffile{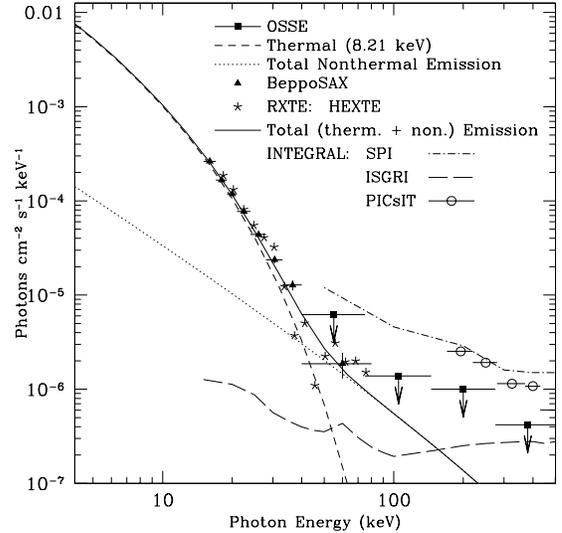}}
\caption{Comparison of X-ray observations of the Coma cluster with the cluster
merger-shock model.  Data points from different instruments are labeled in the
diagram.  The OSSE data points are $2\sigma$ upper limits.  The solid curve is
the sum of nonthermal emission (dotted curve), dominated by Compton-scattered
CMB radiation from primary electrons, and thermal ($\langle T_{\rm X} \rangle
= 8.21\kev$) bremsstrahlung emission (short-dashed curve). Point-source
sensitivities for a 3$\sigma$ detection in $10^6$ seconds with {\it INTEGRAL}
are plotted, though the actual sensitivity is a factor 2 -- 3 worse because of
Coma's extent. }
\label{fig:coma_xray}
\end{center}
\end{figure}

Our model assumes a single merger event. Because galaxy clusters are built up
by multiple merger events \citep{gabici:03}, and GeV -- PeV protons and ions
do not escape from galaxy clusters during a Hubble time
\citep{volk:96,berezinsky:97}, a nonthermal population of protons accelerated
in previous merger events is likely. Coma may have seen $\sim 10$ merger
events during its buildup. Previous merger events are less energetic and
nonthermal protons experience collisional and catastrophic losses through
interactions with the thermal protons in the ICM in the meantime, so the total
energy enhancement of hadrons over leptons is probably not more that $\approx
2$ -- 10 from past merger events. The superposition of proton spectra from
these past events may also produce non-power law spectra \citep{gabici:03}.
This would also hold for the varying shock compression ratios in a single
merger event, though we find that because most of the nonthermal hadrons are
injected when the spectrum is hardest, the proton spectrum is accurately
described by a power-law (Fig.\ \ref{fig1}) with index close to the hardest
reached during the merger event. As demonstrated in the Appendix, injection
indices reaching $\approx 2.2$ are feasible during a minor merger,

The calculated thermal and non-thermal X-ray fluxes are compared in Fig.\
\ref{fig:coma_xray} with the HXR flux measured from the central ($\sim 2.2$
Mpc) region of the Coma cluster.  Data points show the HXR flux measured with
the Phoswich Detection System on {\it Beppo-SAX} \citep{fusco-femiano:04}, the
HEXTE (High-Energy X-Ray Timing Experiment) instrument on {\it RXTE}
\citep{rephaeli:02}, and the OSSE $2\sigma$ upper limits \citep{rephaeli:94}.
Despite the difference in spectral range and field-of-view of the Proportional
Counter Array on {\it Beppo-SAX} and HEXTE on {\it RXTE}, both instruments
indicate the presence of an HXR excess observed in Coma.  The existence of the
HXR tail in the Coma cluster has, however, been challenged
\citep{rossetti:04}. Observations with ISGRI on {\it INTEGRAL} could help to
resolve this controversy, as indicated by the sensitivity curves plotted in
Fig.\ \ref{fig:coma_xray}. But because the Coma Cluster is extended, the
sensitivity is degraded by a factor of 2 -- 3 by the use of a coded-mask
imager on the {\it INTEGRAL} satellite (A.\ Vikhlinin, private communiation,
2004). Astro E-2 (Suzaku) should resolve this important question in the near
future.

\begin{figure}[t]
\begin{center}
\leavevmode
\hbox{%
\epsfxsize=3.0in
\epsffile{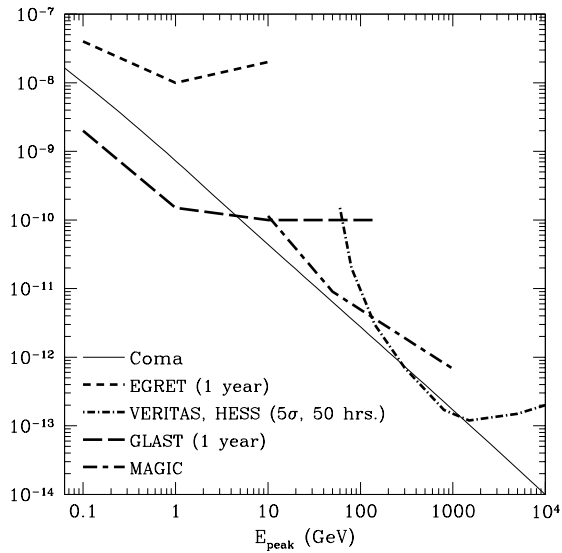}}
\caption{Predicted $\gamma$-ray emission from the Coma cluster of galaxies
from the cluster merger shock model. The solid curve is the predicted photon
flux (in units of ph($> E_{\rm peak}$) cm$^{-2}$ s$^{-1}$).  Sensitivity
limits for EGRET, {\it MAGIC}, {\it GLAST}, and {\it VERITAS} and {\it HESS}
are included.  The EGRET limits are for 2 weeks in the pointing mode, and the
{\it GLAST} limits are for 1 year in its scanning mode.  The quoted {\it
VERITAS}, {\it MAGIC} and {\it HESS} point-source sensitivities are for 50
hour, $5\sigma$ observations \citep{weekes:02}, and will be degraded due to
the angular extent of the Coma cluster's emission. }
\label{fig:coma_highe}
\end{center}
\end{figure}

Fig.\ \ref{fig:coma_highe} shows the predicted $\gamma$-ray emission from the
Coma cluster of galaxies. Sensitivity limits are taken from \citet{weekes:02}
for the {\it VERITAS} point-source sensitivty (which is comparable to the {\it
HESS} sensitivity).  Because of Coma's extent, the sensitivities of these
IACTs can be $\approx 4\times$ worse at 1 TeV \citep{gabici:04} than shown.
The predicted $\gamma$-ray emission falls below the EGRET sensitivity curve
and the measured 2$\sigma$ upper limit of $3.81\times 10^{-8}$ ph($>100$ MeV)
cm$^{-2}$ s$^{-1}$ \citep{reimer:03}.  Our model predicts that {\it GLAST}
will significantly detect the non-thermal $\gamma$-rays from Coma to energies
of several GeV. Reduction in the sensitivity of {\it GLAST} due to the Coma's
extended structure is not significant, as shown in the Appendix.  Furthermore,
we predict that both {\it VERITAS} and {\it HESS} could have high confidence
($\gtrsim 5\sigma$) detections if $\eta_p = 10$\% or there is significant
nonthermal protons left over from past merger events.  Attenuation of sub-TeV
$\gamma$-rays by the extragalactic diffuse infrared and optical radiation
fields is not significant, though it is for multi-TeV fluxes
\citep{gabici:04}.

\section{Discussion}

We have developed a cluster merger model based on observations of merger
activity in the central regions of the Coma cluster of galaxies. Our model
employs a semi-analytic treatment of cluster dynamics based on an N-body
simulation \citep{berrington:03} where nonthermal electrons and protons are
injected with an intensity and spectrum that depends on the strength of the
evolving shocks that are formed by the cluster merger. This represents a
significant improvement over previous analytic merger models
\citep[e.g.,][]{blasi:99,blasi:01,reimer:04} by considering the evolution of
the particle distribution and system parameters with time. The model also
provides a joint treatment of both electron and hadron populations and
detailed spectral fits in comparison with those yet available with numerical
hydrodynamic codes, such as given by \citet{miniati:01} and
\citet{miniati:03}. Moreover, our model is specifically tied to the observed
substructure of Coma.  We find that a detailed merger model gives a good fit
to the multifrequency spectrum of the Coma cluster, confirming the simpler
analytic models.

The measured spectrum of the diffuse radio emission of Coma C is consistent
with a model of synchrotron radiation from electrons accelerated at merger
shocks. Because of the short radiative lifetimes ($\approx 50$ -- 200 Myrs,
depending on magnetic field) for GHz-emitting electrons, compared to the $\sim
0.5$ Gyr merger time scale (see Appendix), the primary synchrotron emission
emitted over the lifetime of the injected electrons is very soft, thus
accounting for the soft radio spectrum shown in Fig.\ \ref{fig:coma_radio}.

The radio emission from Coma C shows an increasingly soft spectral index with
radius \citep{schlickeiser:87,giovannini:93}.  We cannot directly compare our
results with the projected angular distribution of radio emission resulting
from the merger event, because our numerical code does not calculate the
spatial dependence of the radiation emitted by particles.  A merger model may
be expected to produce softer radio emission at large distances from the
cluster center, provided that the contribution of electrons and positrons
formed in secondary nuclear production from protons and ion interactions is
small \citep{miniati:01}.  This is because the speed of the merger shock
varies according to the density contrast between the dominant and smaller
merging cluster.  Because the core and maximum radii, $r_c$ and $r_{\rm max}$,
of the minor cluster are less than that of the dominant cluster (Table 1), the
speed of the shocked fluid becomes monotonically smaller at larger distances
from the collision axis for a planar surface, essentially implying that a bow
shock structure is created (see Appendix). This behavior is already apparent
in Fig.\ 2 of \citep{berrington:03}, where we found that shock strength is
weaker at larger distances from cluster center, so that particles are injected
with softer spectra in the outer parts of the merger, and even more so if they
are away from the merger axis.  Hydrodynamic simulations are needed to
calculate the injection spectra at the same observing times, but it is clear
from our results that softer emission will be injected at larger impact
parameters while the merger event is in progress.

In the case of Coma, excess gas located at a projected radial distance of
$\sim 0.5\mpc$ from the cluster center, if interpreted as gas stripped from
the merging cluster and left at the insertion point on the cluster periphery
($R \approx 2\mpc$), indicates that the angle between the plane of the sky and
the orbital axis of the merger event is $\ltorder 15^{\circ}$
\citep{vikhlimin:97,arnaud:01}. Thus the propagation of the merger shocks are
nearly along the line of sight, and the observed nonthermal emission results
from the line-of-sight integration of the particle emissivities. The lower
density in the periphery of the minor cluster as it merges with the larger
cluster produces, in a planar geometry, stronger reverse shock emission with
potentially harder injection at spectra large radii. If the two clusters were
merging in a direction transverse to the line-of-sight, then we would expect
asymmetric emission related to the merger geometry, as found in off-axis
cluster-merger events \citep{ricker:01}.

A bow shock structure, though with a rather low Mach number (${\cal M} \cong
2.1$), is seen in the merging galaxy cluster A520 \citep{mar05} that displays
a prominent radio halo. Another is found in 1E 0657-56, which also shows a
radio halo \citep{govoni:04}.  This demonstrates that ongoing mergers are
effective at accelerating primary electrons, though additional processes, such
as turbulence, may contribute to acceleration in the radio halo.
 
In the picture studied here, the appearance of Coma is a result of an on-axis
minor merger event where observations are taken shortly after the passage of
the centers of mass of the two merging clusters. This collision epoch is in
fact indicated by the optical data, but could raise questions of whether
observations at this epoch are the result of fine-tuning. It seems likely that
the merger shock is rapidly quenched after the passage and coalescence of the
dark matter halos, so that radio emission is most likely seen precisely at the
epoch at which we find the Coma cluster. This interpretation would also
explain the relative rarity of radio halos in clusters, which would be
primarily limited to those which have undergone recent minor mergers.

The symmetrical and unpolarized radio emissions of most cluster halos may be
contrary to expectations from a merger event, which might be expected to be
highly asymmetric and polarized as for radio relics.  But for the minor
mergers, which are the only ones with sufficient Mach speeds to make radio
halos, a bow-shock ``jellyfish" structure could be created that might be
confused with a quasi-symmetrical radio halo. Viewing through both sides of
the jellyfish could reduce the polarization.  Moreover, high resolution
mapping of the structure of a radio halo could be unfolded to reveal the
extended gaseous and possibly dark matter distributions.  This prospect
becomes more attractive with the development of high resolution, low frequency
radio arrays to map the galaxy cluster radio halo emission.

The difficulty for the cluster merger model is to avoid producing asymmetric
radio emissions, especially for those events observed transverse to the
collision axis.  This problem cannot be alleviated by appealing to MHD
turbulence that moves with the Alfv\'en speed
\begin{equation}
v_{\rm A} = {B\over \sqrt{4\pi \rho}}\cong 7 \; 
{B_{-7} \over \sqrt{n_{-3}}}\;{\rm ~km~s}^{-1}\;
\label{alfvenspeed}
\end{equation}
for $0.1B_{-7}\mu{\rm G~}$ mean magnetic fields and $10^{-3}n_{-3}$ cm$^{-3}$
mean proton densities, which is much slower than the thermal speed. Unless
particles are transported without streaming losses, in which case they do not
deposit energy, energy is transported to the surrounding regions of the merger
shock through MHD processes on timescales much longer than the merging
timescale.  A delayed re-acceleration model can, in principle, remedy the
morphology problem.

To model the radio halo emission from clusters of galaxies,
\citet{schlickeiser:87}, \citet{brunetti:01} and \citet{brunetti:04} consider
reacceleration of electrons via turbulence generated in a post-merger
event. This approach may be supported observationally by detection of
turbulence in the core of Coma \citep{schuecker:04}.  The total energy found
in a population of relativistic nonthermal hadrons is expected to be $\sim 5
\%$ of the thermal energy pool \citep{gabici:03,berrington:03}, inasmuch as
most clusters are expected to experience several merger events in their
formation history and protons with energies $\ltorder 10^{16}\ev$ are confined
within a cluster of galaxies over a Hubble time \citep{volk:96,berezinsky:97}.
This nonthermal pool of streaming protons could introduce turbulence that
would accelerate electrons long after the merger events. Delayed
reacceleration models are highly sensitive to the energy deposited into the
nonthermal protons, however, so that if this energy content is unreasonably
large, an episodic injection scenario, such as the merger-shock model
considered here, may be favored \citep{petrosian:01}.

The morphology question and the discrepant magnetic field values from Faraday
rotation measurements compared to Comptonized CMBR/X-ray emission model values
of $B \cong 0.2 \mu$G \citep{blasi:01,petrosian:01} can in principle be
reconciled with the delayed turbulence reacceleration model
\citep{schlickeiser:87,blasi:01,brunetti:01}. This model has its own
difficulties, however. Second-order Fermi processes accelerate---and very
slowly in the cluster merger environment---hard particle spectra.  A
pre-injection spectrum is required, such as pre-existing particles in the ICM.
Particle spectra with the appropriate spectral index can explain the radio
emission of the Coma cluster in a reacceleration model \citep{brunetti:04},
but have not yet been shown to explain the multiwavelength spectrum, including
the radio spectral softening with radius, by the same model.  Pre-injection of
nonthermal particles from AGN sources \citep{giovannini:93} would also
introduce asymmetries that would have to be ``washed out" in the turbuelent
reacceleration model. Indications that the appearance of radio halos are
correlated with recent cluster merger behavior \citep{buote:01} are, however,
apparently consistent with expectations of the turbulence model \citep{cb05}
given assumptions about the energy density of turbulence with cluster mass.
More detailed studies correlating peaks in radio halo emission and spectra
with merging activity could distinguish between a cluster merger and delayed
reacceleration model but, most likely, both merger shocks and turbulent
reacceleration make important contributions to nonthermal activity in clusters
of galaxies.

Although the halo radio emission from Coma can be explained in the context of
the cluster merger model, we do not claim that all radio halos can be
accounted for by cluster merger shocks.  A striking example that radio halos
may have multiple origins is the recent {\em XMM-Newton} observation of the
galaxy clusters A399 and A401.  Both clusters have radio halos, but only the
halo found in A399 appears to be associated with a shock front resulting from
a cluster merger event \citep{sakelliou:04}.  The origin of nonthermal
electrons producing the radio halo associated with A401 could have a different
acceleration mechanism, for example, arising from accretion shocks
\citep{ensslin:98} or revival of relic non-thermal particles
\citep{ensslin:01}.

Of central importance is to establish the strength and spectrum of the hard
X-ray power-law component reported with {\it BeppoSAX} and {\it RXTE}, which
normalizes the primary nonthermal electron power. Until these reports are
confirmed, which may have to wait for the upcoming launch of Astro E-2,
predictions for $\gamma$-ray detection of clusters of galaxies will remain
highly uncertain.  Moreover, the relative efficiency to accelerate protons and
electrons and the unknown fraction of residual nonthermal hadronic energy left
from previous mergers add additional uncertainties to predictions for
$\gamma$-ray detection.  First-order Fermi acceleration is, however, expected
to accelerate protons with greater efficiency than electrons because of the
larger proton gyro-radii \citep{baring:99}, and past merger events can only
add nonthermal protons to the total.  Thus our $\gamma$-ray predictions are
expected to represent a lower range of the hadronic fluxes, provided that
nonthermal hard X-ray emission has indeed been detected with {\it BeppoSAX}
and {\it RXTE}. If correct, then we predict that the Coma cluster will be
detected with $\approx 5\sigma$ significance with {\it GLAST} (see Appendix),
and marginally or significantly detected with {\it VERITAS} and {\it HESS}.

Our results can be compared with a recent theoretical study \citep{reimer:04}
which concludes that Coma will be detected with {\it GLAST} but not with the
IACTs .  The level of the high-energy radiation fluxes, which is made
primarily by $\pi^0$ decay emission, depends crucially on the spectrum and
high-energy cutoff of the nonthermal hadrons \citep{atoyan:00}, which we argue
is correctly calculated here. Observations of $\gamma$-ray emission from Coma
will reveal the relative efficiency of nonthermal proton and electron
acceleration.

High resolution radio maps from low-frequency ($\lesssim 100 $ MHz) arrays
will be important to determine the nature of the observed radio emission in
Coma.  In the cluster merger scenario, energetic protons will diffuse farther
from their point of origin than lower energy protons, and these have had a
much longer time than the merger timescale to diffuse outwards. At
sufficiently large radii, there will be a transition from radio emission of
primary to secondary origin. Thus the cluster merger model indicates that the
spectral index of Coma's radio halo softens with radius until the radio
emission is dominated by lepton secondaries from hadrons, where it would begin
to harden. A hardening in the volume-integrated mm-radio spectrum of Coma is
likewise expected from a hadronic component.

The $\gtrsim$ TeV $\gamma$-rays from Coma will also be attenuated by the
diffuse extragalactic infrared radiation \citep{gabici:04}.  Our numerical
results from cluster merger model indicate that acceleration of hadronic
cosmic rays by cluster merger shocks will produce a very high-energy
$\gamma$-ray emission component extending to $\gtrsim 10^{16}$ eV. Compton
cascading and reprocessing of the cluster halo ultra high-energy (VHE)
$\gamma$-rays will, for certain values of the intracluster magnetic fields,
produce extended cascade $\gamma$-ray halos. Detection and mapping of such
halos would be an important probe to measure ICM magnetic fields.

\acknowledgments{We thank J.\ Buckley, R. Mushotzky, O.\ Reimer, Y.\ Rephaeli,
C.\ L. Sarazin, and A.\ Vikhlinin for discussion, G.\ Brunetti for helpful
correspondence, and the referee(s) for challenging reports. The research
activity of RB is funded in part by NASA EPSCoR Grant \# NCC5-578.  The
research of RB and CD was also funded by NASA Astrophysical Theory Grant DPR
S-13756-G and NASA {\it GLAST} Science Investigation Grant DPR S-15634-Y. The
work of CDD is supported by the Office of Naval Research.}

\appendix

\section{Analytic Cluster Merger Physics}

\subsection{Radiation Processes}

Suppose that the ICM of Coma is characterized as a quasi-spherical hot cloud
of gas and plasma entraining magnetic field and nonthermal particles that were
injected by shocks formed by merger events during structure formation.  Assume
an isotropic nonthermal electron pitch-angle distribution.  On the size scale
of the telescope's resolution, we also assume that the magnetic field
direction is randomly ordered, though with a mean magnetic field intensity $B$
that is roughly constant through the cluster.

An electron with Lorentz factor $\gamma$ emits photons with frequency
$\nu_{\rm syn}($Hz$)\cong 2.8\times 10^6 B({\rm G})\gamma^2$.  For magnetic
fields measured in units of $10^{-7} B_{-7}$ G, and for radio measurements at
$\nu_{\rm GHz}$ GHz, electrons with Lorentz factor
\begin{equation}
\gamma_{\rm syn} \cong 6\times 10^4 \sqrt{{\nu_{\rm GHz}\over B_{-7}}}\;,
\label{gammasyn}
\end{equation}
Electrons with Lorentz factor
\begin{equation}
\gamma_{\rm C} \cong 2.5\times 10^4 \sqrt{{\epsilon\over 1+z}}\;,
\label{gammaC}
\end{equation}
Thomson-scatter CMBR photons to X-ray energies ($\epsilon =0.1$ corresponds to
a photon of $\approx 50$ keV).

Let $\bar \nu_{\rm GHz}$ and $\bar \epsilon$ denote the radio frequency and
dimensionless hard X-ray photon energy where the spectral slopes of the
nonthermal radiation are the same.  For Coma, we find that a nonthermal photon
flux with photon number spectral index $\cong -1.8$ ($\alpha = 0.8$) provides
a good fit to the {\it BeppoSAX} \citep{fusco-femiano:04} and ({\it RXTE})
\citep{rephaeli:02} observations (Fig.\ \ref{fig:coma_xray}).  The radio index
$\alpha = 0.8$ at 30 MHz energies (Fig.\ \ref{fig:coma_radio}).  The
interpretation that the nonthermal radio emission is synchrotron radiation and
that Coma displays an associated nonthermal hard X-ray component from
Compton-scattered CMBR implies magnetic fields from the condition that the
primary Compton and primary synchrotron fluxes should have the same spectral
indices for electrons with the same energies, that is, $\gamma_{\rm syn} =
\gamma_{\rm C}$. Hence
\begin{equation}
B_{-7} \cong 5.8 \; {\bar \nu_{\rm GHz}\over \bar
\epsilon }\;(1+z)\;.
\label{B-7}
\end{equation}
For Coma, therefore, $B_{-7} \cong 2.3$, in good agreement with the value
$B_{-7} = 2.2$ used in the numerical model. The discrepancy between this value
and values derived from Faraday rotation measurments could be explained by
multiple zones \citep{blasi:01,petrosian:01}, or the attribution of the hard
X-ray excess to nonthermal bremsstrahlung \citep{sarazin:00}. Confirmation of
the {\it Beppo-SAX} and {\it RXTE} results is especially crucial here, because
the magnetic field can be larger and the electron intensity correspondingly
smaller if there is no nonthermal X-ray tail to explain.

The timescale for Thomson-scattered CMBR and synchrotron cooling can be
written as
\begin{equation}
t_{C_s}\cong {2.5\times 10^{6} \over \gamma[ (1+z)^4 + 10^{-3}B_{-7}^2]} 
\myr.
\label{tCs}
\end{equation}
The lifetime for electrons that emit synchrotron radio emission with frequency
$\nG$ GHz is
\begin{equation}
t_{C_s}(\gamma_{\rm syn})\cong {41 \sqrt{B_{-7}/\nG } \over (1+z)^4 +
10^{-3}B_{-7}^2}\myr.
\label{tCss}
\end{equation}
The lifetime for electrons that emit hard X-ray photons with dimensionless
energy $\epsilon = 0.1\epsilon_{-1}$ is
\begin{equation}
t_{C_s}(\gamma_{\rm C})\cong {320 \sqrt{(1+z)/\epsilon_{-1} }\over
(1+z)^4 + 10^{-3}B_{-7}^2}\myr.
\label{tCsx}
\end{equation}
These timescales are generally shorter than dynamical time scales, as we now
show.
 
\subsection{Scaling Relations for Cluster Dynamics}

From elementary considerations, the gain in kinetic energy when the minor
cluster, treated as a test particle in the mass distribution of the dominant
cluster, falls from radius $r_1$ to radius $r_2 (\leq r_1) $, is $M_2 v_2^2/2
= GM_1 M_2 (r_2^{-1} -r_1^{-1}$). Thus $v\cong \sqrt{2GM_1/r_2}$ when $r_2 \ll
r_1$, so that
\begin{equation}
v_2\approx 6000 \;\sqrt{M_{15}\over (r_2/0.257\mpc)}\kms,
\label{vsim}
\end{equation}
where $r_2$ is equated with the core radius of the dominant cluster (see Table
1).  The characteristic merger time $\hat t$ is determined by the acceleration
$a = GM_1/r_1^2$ at the outer radius.  Because $ r_1 \approx a_2 \hat t^2/2$,
\begin{equation}
\hat t \cong \sqrt{{2r_1^3\over GM_1}}\approx 660\; {r^{3/2}_{\rm Mpc}
 \over M_{15}}\;{\rm~Myr}\;,
\label{tcongr}
\end{equation}
where $M_1 = 10^{15} M_\odot$ and $r_1 = r_{\rm Mpc}$ Mpc.  The available
energy in the collision is
\begin{equation}
{\cal E} \approx {GM_1 M_2\over r} \approx {8\times 10^{63}\over r_{\rm
Mpc}}\; M_{15}\; \left({M_2\over 10^{14}M_\odot}\right)\ergs\;.
\label{calE}
\end{equation}
The long timescale of merger events compared to the timescale for electrons
that emit GHz radio emission, eq.\ (\ref{tCss}), means that these electrons
had to have been accelerated recently compared to the duration of the cluster
merger event.

The sound speed 
\begin{equation}
c_s(T_X) =\sqrt{k_{\rm B}T\over \langle m \rangle } \; \cong \; 1200
 \;\sqrt{k_{\rm B} T_x \over 10\kev}\kms,
\label{csTX}
\end{equation}
where the mean mass per particle is taken to be $\langle m \rangle = 0.6 m_p$.
From eq.\ (\ref{vsim}), this implies shocks with moderate Mach numbers ${\cal
M} \approx 5$.

\subsection{Analytical Cluster Dynamics}

We treat cluster dynamics analytically \citep[see also][]{ricker:01}, as a
check on the validity of the numerical treatment. The merging activity in the
Coma cluster of galaxies is again treated in the test-particle approximation
for the minor cluster with mass $M_2$, though now using the $\beta$-model
approximation to the isothermal mass distribution \citep{calaviere:76} given
by
\begin{equation}
\rho(r) = \rho_0 \left[1 + \left(\frac{r}{r_c}\right)^{2} \right]^{-3\beta/2}
\label{eqn:isothermal_beta}
\end{equation}
for the dominant cluster, with $\rho_0 = \zeta \rho_{0,1}$, from eq.\
(\ref{eqn:king_approximation}).  The mass of cluster 1 interior to radius $r$
is
\begin{equation}
M_1(<r) = 4\pi \rho_0 r_c^3 \int_0^{r/r_c}dx\; \frac{x^2}{1 + x^2} \;.
\label{M1(r)}
\end{equation}
We approximate $\beta = 2/3$ for Coma, which is close to its value, $\beta =
0.705$, determined from observations.  The outer radius $r_{\rm max}$ is
normalized to the total mass $M_1$ of the dominant cluster, giving
\begin{equation}
M_1 = 4\pi r_c^3 \left[{r_{\rm max} \over r_c} - \arctan \left({r_{\rm
max}\over r_c} \right) \right]\rho_0 \;.
\label{M1}
\end{equation}
Solving for $\zeta = 21$, corresponding to a ratio of normal mass to CDM mass
of 5\%, gives $r_{\rm max} =$ 0.82 Mpc. When $\zeta = 11$, for a 10\% ratio of
normal mass to CDM mass, $r_{\rm max} =$ 1.3 Mpc.

The change in potential energy of the test-particle cluster 2 when falling
through the gravitational potential of cluster 1 is
\begin{equation}
\Delta U = {1\over 2}M_2 (\Delta v)^2 = -4\pi G \rho_0 M_2 r_c^3 
\int_{r_1}^{r_2} dr\; r^{-2}\;
\int_0^{r/r_c} dy\; \frac{y^2}{(1+y^2)^{3\beta/2}}\;,
\label{DU}
\end{equation}
provided $r_{\rm max} \geq r_1 > r_2$. For the case $\beta = 2/3$, eq.\
(\ref{DU}) can be solved to give
\begin{equation}
(\Delta v)^2 = 4\pi G \rho_0 r_c^2 \left[ \ln \left({1+ y_1^2\over 1+
y_2^2}\right) + 2\left( {\arctan y_1 \over y_1} - {\arctan y_2 \over
y_2}\right)\right]\;,
\label{deltav2}
\end{equation}
where $y_{1(2)} = r_{1(2)}/r_c$. 

Taking $y_2\rightarrow 0$, we obtain $\Delta v = 4760$ and $\Delta v = 3750$
km s$^{-1}$ for $\zeta = 21$ and 11, respectively. A sound speed $c_s \cong
1100$ km s$^{-1}$ of the dominant cluster (Table 1 and eq.\ [\ref{csTX}])
implies Mach numbers ${\cal M} = 4.37$ and $3.44$, compression ratios $\chi =
3.5$ and 3.2, and injection indices $s = 2.2$ and $2.4$ for $\zeta = 21$ and
11, respectively, using the standard formulas for nonrelativistic shock
acceleration in the test particle limit.  The injection index $s = 2.2$ is in
good agreement with the numerical results (see Fig.\ \ref{fig1}).

We instead consider the dark-matter density profile distribution given by a
\begin{equation}
\rho(r) = \rho_s \; \left({\frac{r}{r_s}}\right)^{-\alpha} 
\left[1 + \left(\frac{r}{r_s}\right)^{\gamma} \right]^{-\beta}\;,
\label{NFW}
\end{equation}
where $r_s$ is the scaling radius and $\rho_s$ is the characteristic
overdensity.  For the NFW \citep{navarro:97} model, $(\alpha, \beta, \gamma) =
(1,2,1)$, and the mass of cluster 1 interior to radius $r$ becomes
\begin{equation}
M_1(<r) = 4\pi \rho_s r_s^3 \int_0^{r/r_s}dx\; \frac{x}{(1 + x)^2} \; = \;
4\pi \rho_s r_s^3\;\left[\ln \left(1 + \frac{r}{r_s}\right)\; -\; {\frac{r/r_s}
{1+r/r_s}}\right]\;.
\label{MN(r)}
\end{equation}
By normalizing to the total mass $M_1 = M_1(<r_{\rm max})$ at the outer radius
$r_{\rm max}$, one obtains that $r_{\rm max} = 843$ kpc for $M_1$ from Table
1, with $\rho_s = 1.13\times 10^{-25}$ g cm$^{-3}$ \citep[using $r_s = 459$
kpc and $c = 5.42$ given by][]{ettori:02}. Deriving the speed following the
approach leading to eq.\ (\ref{deltav2}) gives
\begin{equation}
(\Delta v)^2 = 8\pi G \rho_s r_s^2 \left.\left[\frac{\ln
(1+y)}{y}\right]\right|_{r_2/r_s}^{r_1/r_s} \;.
\label{DU1}
\end{equation}
In the limit $r_2 \ll r_s$ and $r_1 \cong r_{\rm max}$, one finds that $\Delta
v \cong 4040$ km s$^{-1}$, implying a Mach number of ${\cal M} = 3.7$,
intermediate to the values found for the $\beta$-model profile.  Thus, our
conclusions should not be much different whether we consider a $\beta$-model
or NFW profile.

\subsection{Shock Structure in Merging Clusters of Galaxies}

In \cite{berrington:03}, we derived the Mach numbers ${\cal M}_1$ and ${\cal
M}_2$ of the forward and reverse shocks, respectively, given by
\begin{equation}
{\cal M}_1(t) = \frac{2} {3} \frac{v} {c_{1}} \left( 1 + \sqrt{1 + \frac{9}
{4} \frac{c_{1}^{2}} {v^{2}} } \right) \;,
\label{forward}
\end{equation}
and
\begin{equation}
{\cal M}_2(t) = \frac{2} {3} \frac{v_{0}(t) - v} {c_{2}} \left( 1 + \sqrt{1 +
\frac{9} {4} \frac{c_{2}^{2}} {(v_{0}(t) - v)^{2}} } \right)\;.
\label{reverse}
\end{equation}
Here $c_{1}$ and $c_2$ are the sound speeds in the dominant and merging
clusters, respectively.  The value of $v$ is calculated by iteratively solving
\begin{equation}
\frac{\mu_{1}}{\mu_{2}} \frac{n_{1}(t)}{n_{2}(t)} = \frac{1 + 3{\cal
    M}_{1}^{-2}(t)}{1 + 3 {\cal M}_{2}^{-2}(t)} 
	\left( \frac{v_{0}(t) - v}{v}
    \right)^{2}\;,
\label{reverse1}
\end{equation}
using eqs.\ (\ref{forward}) and (\ref{reverse}).  The terms $n_{1}(t)$ and
$n_{2}(t)$ are the ICM number densities given by eq.\
(\ref{eqn:isothermal_beta}), and $\mu_1$ and $\mu_2$ are the mean masses per
particle in the dominant and merging clusters, respectively.

A core radius of 0.15 Mpc is used for the minor cluster in the
calculations. Becasue the core radius of the minor cluster is smaller than
that of the dominant cluster, the relative densities along the shock front
decrease away from the collision axis joining the two merging clusters. The
shocked fluid speed also monotonically decreases away from the collision axis,
as can be seen by numerically solving the above equations. This produces a
weaker forward shock and a bow shock structure that will inject softer
electron spectra at greater angular distances from the center of the Coma
cluster.

\begin{figure}[t]
\begin{center}
\leavevmode
\hbox{%
\epsfxsize=5.0in
\epsffile{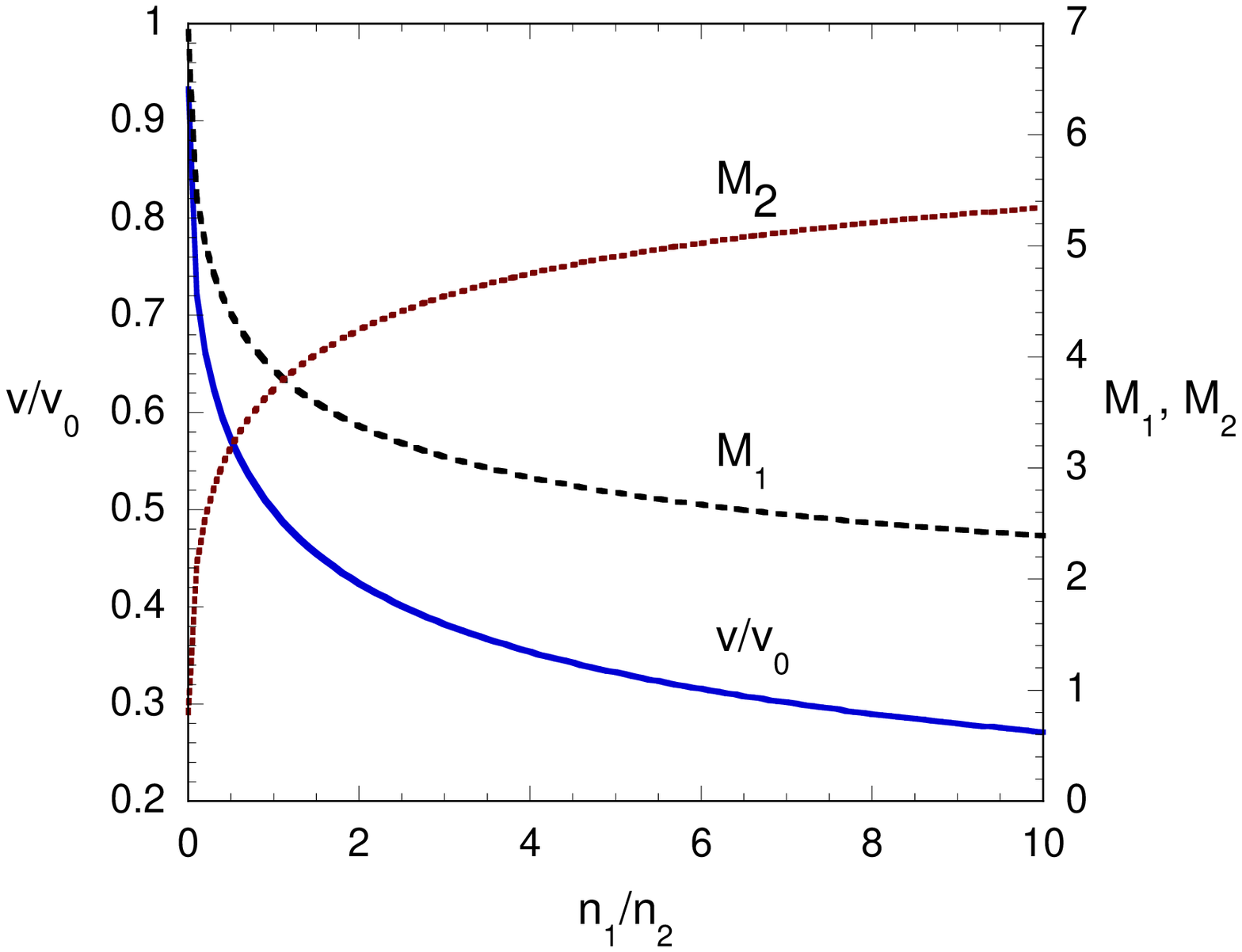}}
\caption{The ratio $v/v_0$ of the shocked fluid speed to the merger speed
$v_0$, and the Mach numbers ${\cal M}_1$ and ${\cal M}_2$ of the forward and
reverse shocks, respectively, are plotted as a function of the ratio of the
densities of the dominant and merging clusters.  }
\label{fig7}
\end{center}
\end{figure}

The speed $v$ of the shocked fluid and the Mach numbers of the forward and
reverse shocks calculated from these equations are shown in Fig.\ \ref{fig7}
as a function of density contrast $n_1/n_2$ in a planar geometry. The relative
speed $v_0$ of the two clusters is set equal to 6000 km s$^{-1}$.  The gas
temperatures of the two clusters are given in Table 1.  When $n_1/n_2$ is
larger, the speed of the shocked fluid is smaller.  Consequently a bow-shock
structure is formed.

\subsection{Detection of Coma with {\it GLAST}}

Using the expressions derived in Section 4 of \citet{berrington:03}, we
estimate the detection significance and number of photons that GLAST will
detect from the Coma cluster during $t_{yr}$ years of GLAST observations in
the scanning mode, using an exposure factor for Coma of $20$\%.

Figs.\ \ref{fig2} and \ref{fig3} show that the $\gamma$-ray flux spectrum from
Coma can be expressed as
\begin{equation}
\nu L_\nu = 10^{43} L_{43} \left( {E_\gamma\over
100{\rm~MeV}}\right)^{\alpha_\nu}\;{\rm ergs~s}^{-1}\;,
\label{nuLnu}
\end{equation}
where $\alpha_\nu = -0.2$ and $L_{43} \lesssim 0.3$ (in the best case still
allowed by the high-frequency radio data, $\eta_p \cong 20$\% and $L_{43}
\approx 0.6$).

The number of photons with energies $\gtrsim E_\gamma$ detected from the Coma
cluster with GLAST is
\begin{equation}
N_s (>E_\gamma ) \cong 35 \;t_{yr} [E_\gamma({\rm GeV})]^{-1.04}\;\left(
{L_{43}\over 0.3}\right)\;.
\label{Ns}
\end{equation}
More photons are detected at lower energies, but this expression is only valid
above the $\pi^0\rightarrow 2\gamma$ bump, which is found at a few hundred MeV
in a $\nu F_\nu$ representation.

The best detection significance is found by assuming that the emission from
Coma is point-like, which holds until the point-spread function of GLAST is
smaller than the angular extent of Coma. The Coma C radio source has an
angular extent of $\approx 20^\prime$ \citep{giovannini:93}, which is
comparable to the psf of $\approx 2$ GeV photons. The $n_\sigma$ detection
significance for Coma is found to be
\begin{equation}
n_\sigma \cong 5.4\;\sqrt{t_{yr}}\;\left( {L_{43}\over
  0.3}\right)\;[E_\gamma({\rm GeV})]^{0.1}\;.
\label{nsigma}
\end{equation}
These two expressions show that Coma will be significantly detected with {\it
GLAST} if the energy injected in nonthermal hadrons is $\approx 10\times$ the
energy in nonthermal electrons (or if there is significant energy in
nonthermal protons left over from previous merger events). The number of
detected photons, which may be as large as a few hundred, would permit
spectral analysis to determine if the spectrum is consistent with a $-2.2$
photon spectral index at GeV energies. If the nonthermal hadron energy content
has been overestimated, however, it remains possible that that GLAST will only
weakly detect the Coma cluster of galaxies.

\end{document}